\documentclass[aps,twocolumn,showpacs]{revtex4}
\usepackage{graphicx}

\begin{document}

\title{Application of the diffusion equation to prove scaling invariance on the 
transition from limited to unlimited diffusion}

\author{$^1$Edson D.\ Leonel, $^1$C\'elia Mayumi Kuwana, $^1$Makoto Yoshida, 
$^2$Juliano Antonio de Oliveira}

\address{$^1$Universidade Estadual Paulista (UNESP) - Departamento de 
F\'isica\\ Av.24A, 1515 -- Bela Vista -- CEP: 13506-900 -- Rio Claro -- SP -- 
Brazil\\
$^2$Universidade Estadual Paulista (UNESP) - Campus de S\~ao Jo\~ao da 
Boa Vista\\ Av. Prof$^a$. Isette Corr\^ea Font\~ao, 505 -- CEP: 13876-750 -- 
S\~ao Jo\~ao da Boa Vista -- SP -- Brazil}

\date{\today} \widetext


\begin{abstract}
The scaling invariance for chaotic orbits near a transition from unlimited to 
limited diffusion in a dissipative standard mapping is explained via the 
analytical solution of the diffusion equation. It gives the probability of 
observing a particle with a specific action at a given time. We show the 
diffusion coefficient varies slowly with the time 
and is responsible to suppress the unlimited diffusion. The momenta of the 
probability are determined and the behavior of the average squared action is 
obtained. The limits of small and large time recover the results known in the 
literature from the phenomenological approach and, as a bonus, a scaling for 
intermediate time is obtained as dependent on the initial action. The formalism 
presented is robust enough and can be applied in a variety of other systems 
including time dependent billiards near a transition from limited to unlimited 
Fermi acceleration as we show at the end of the letter and in many other 
systems under the presence of dissipation as well as near a transition from 
integrability to non integrability.
\end{abstract}

\maketitle

Since the observation and a phenomenological characterization \cite{Paper1} of
scaling invariance in the chaotic sea near a transition from integrability to 
non integrability in a Fermi Ulam model \cite{Paper2}, the formalism using 
homogeneous and generalized function leading to a set of critical exponents 
\cite{Book1} has been widely used in a variety of systems to investigate 
dynamical properties near dynamical phase transitions including oscillating 
spring mass system \cite{Paper3}, billiards 
\cite{Paper4,Paper5}, scaling in social media \cite{Paper6}, in waveguides 
\cite{Paperadd} and in many other systems. In a majority of the cases the 
scaling is closely connected with diffusion yielding applications in different 
subjects of science being therefore observed in systems from pollen diffusing 
\cite{Paper7}, in disease propagation \cite{Paper8,Paper9,Paper10}, in pests 
spreading \cite{Paper11} and many others hence making the topic of wide 
interest.

In this letter our aim is to characterize analytically a transition from limited 
to unlimited diffusion by using the diffusion equation \cite{Book2} applied in a 
paradigmatic model in nonlinear science the so called dissipative standard 
mapping \cite{Book3} and the success of the formalism allowed us to extend 
its applicability to time dependent billiards. The mapping is written in terms 
of two equations $I_{n+1}=(1-\gamma)I_n+\epsilon\cos(\theta_n)$ and 
$\theta_{n+1}=(\theta_n+I_{n+1})~\rm{mod}(2\pi)$ where $\gamma\in[0,1]$ is the 
dissipative parameter and $\epsilon$ corresponds to the intensity of the 
nonlinearity. This system has two well known transitions \cite{Book3} for 
$\gamma=0$ (conservative case): (i) A transition from integrability for 
$\epsilon=0$ where the phase space is foliated to non integrability when 
$\epsilon\ne0$ and mixed structure is presented in the phase space including 
periodic islands, chaotic seas and invariant spanning curves limiting the 
diffusion to a closed region; (ii) at a critical value of 
$\epsilon_c=0.9716\ldots$, the system admits a transition from local chaos when 
$\epsilon<\epsilon_c$ to globally chaotic dynamics for $\epsilon>\epsilon_c$ 
where invariant spanning curves are no longer present and, depending on the 
initial conditions, chaos can diffuse unbounded in the phase space. The 
determinant of the Jacobian matrix is $\det~J=(1-\gamma)$ and for $\gamma\ne0$ 
the Liouville's theorem is violated leading to the existence of attractors in 
the phase space. For large enough $\epsilon$, typically $\epsilon>10$ sinks are 
not observed in the phase space hence leading the dynamics to have chaotic 
attractors in the limit of small values of $\gamma$. At such limit one is 
facing a transition from limited ($\gamma\ne0$) to unlimited 
($\gamma=0$) diffusion for the variable $I$ which is the transition we consider 
in this letter. Our main goal in this letter is to fix up an open problem in the 
nonlinear community discussing the scaling invariance present in the transition 
from limited for $\gamma\ne0$ to unlimited diffusion when $\gamma=0$, so far 
analytically for large values of $\epsilon$. As far as we can tell, this scaling 
investigation has only been described using a phenomenological approach 
\cite{Paper12} assuming a set of scaling hypotheses allied with a homogeneous 
function hence leading to a set of critical exponents leaving a lack on the 
analytical solution which to the best knowledge of the authors has never been 
made. At the same time, this letter fix up this gap in the literature and the 
present approach is proved to be valid and can be used in a wide class of other 
systems including transition from limited to unlimited Fermi acceleration in 
time dependent billiards as we shown in the end of the letter, integrability to 
non integrability in nonlinear mappings and many others.

\begin{figure}[t]
\centerline{\includegraphics[width=1\linewidth]{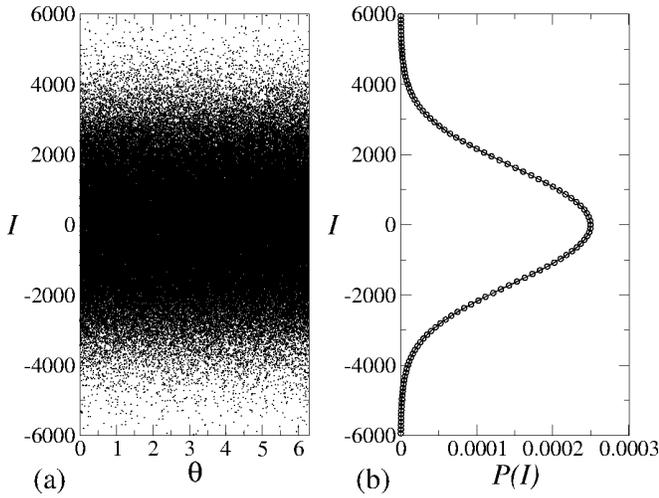}}
\caption{(a) Plot of the phase space for the standard dissipative mapping 
considering the parameters $\epsilon=100$ and $\gamma=10^{-3}$. (b) Normalized 
probability distribution for the chaotic attractor shown in (a).}
\label{Fig1}
\end{figure}

The range of parameters we are interested in to validate the transition is 
$\gamma$ positive and small, typically $\gamma\in[10^{-5},10^{-2}]$ and 
$\epsilon>10$, which drives the system to high nonlinearities and absence of 
sinks in the phase space. At such a window of parameters a transition from 
limited, $\gamma\ne0$ to unlimited, $\gamma=0$, diffusion is observed. A 
typical plot of the phase space is shown in Figure \ref{Fig1}(a) 
illustrating a chaotic attractor for the parameter $\epsilon=10$ and 
$\gamma=10^{-3}$ together with the probability distribution along the chaotic 
attractor shown in Figure \ref{Fig1}(b). We see from Figure \ref{Fig1}(a) the 
density of points concentrate around $I\cong0$ and is symmetric with respect to 
the vertical axis. The distribution fades soon as it goes apart from the 
origin. The positive Lyapunov exponent measured \cite{Paper13} for the chaotic 
attractor shown in Fig. \ref{Fig1}(a) was $\lambda=3.9120(1)$.

Given an initial condition near $I\cong0$ the particle diffuses along the 
chaotic attractor. The natural observable to characterize the diffusion is the 
average squared action $I_{rms}(n)=\sqrt{{{1}\over{M}}\sum_{i=1}^MI^2_i}$ where 
$M$ corresponds to an ensemble of different initial conditions along the chaotic 
attractor. To obtain such observable we need to solve the diffusion equation 
that gives the probability to observe a specific action $I$ at a given time $n$, 
i.e. $P(I,n)$. The diffusion equation is written as 
\begin{equation}
{{\partial P(I,n)}\over{\partial n}}=D{{\partial^2P(I,n)}\over{\partial I^2}},
\label{eq1}
\end{equation}
where the diffusion coefficient $D$ is obtained from the first equation of 
the mapping by using $D={{\overline{I^2}_{n+1}-\overline{I^2_n}}\over{2}}$. 
A straightforward calculation assuming statistical independence between $I_{n}$ 
and $\theta_n$ at the chaotic domain leads to
\begin{equation}
D(\gamma,\epsilon,n)={{\gamma(\gamma-2)}\over{2}}\overline{I^2}_n+{{\epsilon^2}
\over{4}}.
\label{eq2}
\end{equation}
The expression of $\overline{I^2}_n$ is obtained also from the first equation 
of the mapping assuming that 
$\overline{I^2}_{n+1}-\overline{I^2}_n={{\overline{I^2}_{n+1}-\overline{I^2}_n} 
\over{(n+1)-n}}\cong{{d\overline{I^2}}\over{dn}}=\gamma(\gamma-2)\overline{I^2}+
{{\epsilon^2}\over{2}}$, whose solution is
\begin{equation}
\overline{I^2}(n)={{\epsilon^2}\over{2\gamma(2-\gamma)}}+\left(I_0^2+{{
\epsilon^2 } \over { 2\gamma(\gamma-2)}}\right)e^{-\gamma(2-\gamma)n}.
\label{eq3}
\end{equation}
To compare with the experimental observable Eq. (\ref{eq3}) must be averaged 
over the orbit, leading to
\begin{eqnarray}
<&\overline{I^2}&(n)>={{1}\over{n+1}}\sum_{i=0}^n\overline{I^2}(i)={{
\gamma(\gamma-2)}\over{2(n+1)}}\times\nonumber\\
&\times&\left[I^2_0+{{\epsilon^2}
\over { 2\gamma(\gamma-2)}}\left({{1-e^{-(n+1)\gamma(2-\gamma)}}\over{1-e^{
-\gamma(2-\gamma) } } } \right)\right].
\label{eq4}
\end{eqnarray}

To obtain an unique solution for Eq. (\ref{eq1}) we impose the following 
boundary conditions $\lim_{I\rightarrow \pm\infty}P(I)=0$ with the initial 
condition $P(I,0)=\delta(I-I_0)$ that warrants all particles leave from the 
same initial action but with $M$ different initial phases $\theta\in[0,2\pi]$. 
Although the diffusion coefficient $D$ depends on $n$ its variation is slow 
and little from the instant $n$ to $n+1$. This property 
allows us to consider it {\it constant} to obtain the solution of the diffusion 
equation. However, soon as the solution is obtained, the expression of $D$ from 
Eq. (\ref{eq2}) is incorporated to the solution. The technique used to solve 
Eq. (\ref{eq1}) is the Fourier transform \cite{Book4}. Because the probability 
is normalized, i.e. $\int_{-\infty}^{\infty}P(I,n)dI=1$, we can define a 
function
\begin{equation}
R(k,n)=\mathcal{F}\{P(I,n)\}={{1}\over{\sqrt{2\pi}}}\int_{-\infty}^{\infty}P(I,
n)e^{ikI}dI.
\label{eq5}
\end{equation}
Differentiating $R(k,n)$ with respect to $n$ and from the property that 
$\mathcal{F}\left\{{{\partial^2P}\over{\partial I^2}}\right\}=-k^2R(k,n)$ we 
end up the following equation to be solved ${{dR}\over{dn}}(k,n)=-Dk^2R(k,n)$, 
which leads to
\begin{equation}
R(k,n)=R(k,0)e^{-Dk^2n}.
\label{eq6}
\end{equation}
Considering the initial condition we have that 
$R(k,0)=\mathcal{F}\{\delta(I-I_0)\}={{1}\over{\sqrt{2\pi}}}e^{ikI_0}$. 
Inverting the expression of $R(k,n)$ we obtain
\begin{eqnarray}
P(I,n)&=&{{1}\over{\sqrt{2\pi}}}\int_{-\infty}^{\infty}R(k,n)e^{-ikI}
dk,\nonumber\\
&=&{{1}\over{\sqrt{4\pi Dn}}}e^{-{{(I-I_0)^2}\over{4Dn}}}.
\label{eq7}
\end{eqnarray}
Equation (\ref{eq7}) satisfies both the boundary and initial condition as well 
as the diffusion equation (\ref{eq1}). It is also normalized by construction. 
The observable we want to characterize is 
$\overline{I^2}(n)=\int_{-\infty}^{\infty}I^2P(I,n)dI$, which leads to 
$\overline{I^2}(n)=\sqrt{2D(n)n+I_0^2}$. Using $D(n$) 
obtained from Eq. (\ref{eq2}), we end up with the expression of $I_{rms}(n)$ as
\begin{widetext}
\begin{equation}
I_{rms}(n)=\sqrt{I_0^2+{{n\gamma(\gamma-2)}\over{n+1}}\left[
I_0^2+{{\epsilon^2}\over{2\gamma(\gamma-2)}}\right]\left[{{
-(n+1)\gamma(2-\gamma)}\over{1-e^{-\gamma(2-\gamma)}}}\right]}.
\label{eq8}
\end{equation}
\end{widetext}

Let us discuss specific limits of $n$ and its consequences to Eq. (\ref{eq8}). 
The first limit is $n=0$, which leads to $I_{rms}(0)=I_0$, in well agreement to 
the initial condition. The second limit is $n\rightarrow\infty$. At such a 
limit we have
\begin{equation}
I_{rms}=\sqrt{I_0^2+\gamma(\gamma-2)\left[
I_0^2+{{\epsilon^2}\over{2\gamma(\gamma-2)}}{{1}\over{1-e^{-\gamma(2-\gamma)}}}
\right]},
\label{eq9}
\end{equation}
and that when expanding in Taylor series up to first order the term 
$1-e^{-\gamma(2-\gamma)}\cong\gamma(2-\gamma)$ we obtain 
\begin{equation}
I_{rms}={{1}\over{\sqrt{2(2-\gamma)}}}\epsilon\gamma^{-1/2}.
\label{eq10}
\end{equation}
Let us discuss this result prior move on. It is known in the literature 
\cite{Paper12} that the critical exponents $\alpha_1$ and $\alpha_2$ can be 
obtained from the scaling theory. It was supposed that for large enough $n$, 
the stationary state is given by $I_{rms}\propto 
\epsilon^{\alpha_1}\gamma^{\alpha_2}$. An immediate comparison of this scaling 
hypothesis with Eq. (\ref{eq10}) leads to a remarkable results of $\alpha_1=1$ 
and $\alpha_2=-{{1}\over{2}}$, in very well agreement with the phenomenological 
prediction discussed in Ref. \cite{Paper12}. Interestingly, such a result can 
also be obtained from the own equations of the mapping imposing that 
$\overline{I^2}_{n+1}=\overline{I^2}_n=\overline{I^2}_{sat}$, yielding 
$I_{sat}={{1}\over{\sqrt{2(2-\gamma)}}}\epsilon\gamma^{-1/2}$.

The limit of small $n$ is the third limit we consider. Assuming that the 
initial action $I_0\cong0$, hence negligible as compared to $\epsilon$ and 
doing a Taylor expansion on the exponential of the numerator from Eq. 
(\ref{eq9}) we obtain $I_{rms}(n)\cong\sqrt{{{\epsilon^2}\over{2}}n}$. This 
result proves that for short $n$, an ensemble of particles diffuses along the 
chaotic attractor analogously as a random walk motion, hence with diffusion 
exponent $\beta=1/2$, i.e., normal diffusion. From Ref. \cite{Paper12} a 
scaling hypothesis at the limit of small $n$ is $I_{rms}(n)\propto 
(n\epsilon^2)^{\beta}$, with $\beta=1/2$ in well agreement with the theoretical 
prediction discussed above.

A fourth interesting limit we want to take into account is again intermediate 
$n$ but non negligible $I_0$ such that $0<I_0<I_{sat}$. At such windows of 
$I_0$ and $n$, an additional crossover is observed when 
$n_x^{\prime}\cong2{{I_0^2}\over{\epsilon^2}}$. This crossover had already 
been observed in \cite{Paper1} when a phenomenological approach was proposed 
and confirmed analytically in \cite{Paper14}.

A fifth limit is in the case of $I_0\cong 0$, leading to a growth in $I_{rms}$ 
for short $n$ followed by a crossover and a bend towards the regime of 
saturation. Such a characteristic crossover is given by 
$n_x\cong{{1}\over{2-\gamma}}\gamma^{-1}$. From the scaling approach as 
discussed in Ref. \cite{Paper12} it is assumed that $n_x\propto 
\epsilon^{z_1}\gamma^{z_2}$ and that $z_1=0$ and $z_2=-1$, as obtained above.

The last regime of interest is considered when $I_0\gg 
{{\epsilon^2}\over{2\gamma(2-\gamma)}}$. At this limit, Equation (\ref{eq8}) is 
rewritten as
\begin{equation}
I_{rms}(n)=\sqrt{I_0^2e^{-(n+1)\gamma(2-\gamma)}+\epsilon^2
{{(1-e^{-(n+1)\gamma(2-\gamma)})}\over{2\gamma(2-\gamma)}}}.
\label{eq11}
\end{equation}
The leading term for small $n$ is 
$I_{rms}(n)=I_0e^{-(n+1){{\gamma(2-\gamma)}\over{2}}}$ while the stationary 
state is obtained at the limit of 
$\lim_{n\rightarrow\infty}I_{rms}={{\epsilon}\over{\sqrt{2(2-\gamma)}}}\gamma^{
-1/2 } $ , 
in well agreement with the previous results.

\begin{figure}[b]
\centerline{\includegraphics[width=1\linewidth]{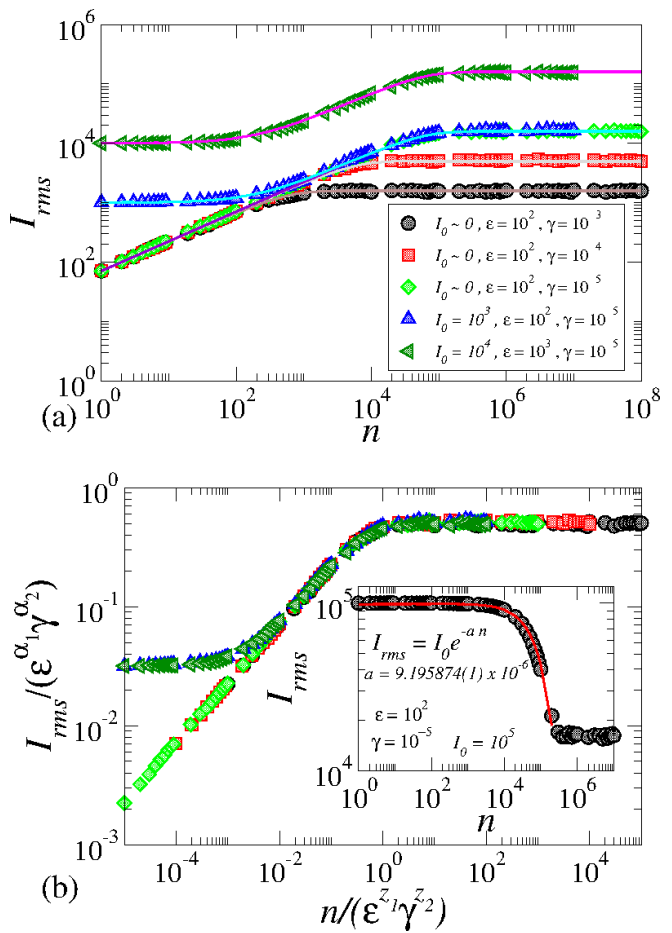}}
\caption{(a) Plot of the phase space for the standard dissipative mapping 
considering the parameters $\epsilon=100$ and $\gamma=10^{-3}$. (b) Normalized 
probability distribution for the chaotic attractor shown in (a). Inset of (b) 
shows an exponential decay to the attractor.}
\label{Fig2}
\end{figure}

Figure \ref{Fig2}(a) shows 
a plot of $I_{rms}~vs.~n$ for different control parameters and initial 
conditions, as labeled in the figure. Filled symbols correspond to the numerical 
simulation obtained direct from the iteration of the dynamical equations of the 
mapping considering an ensemble of $M=10^3$ different initial particles, all 
starting with same action $I_0$, as shown in Fig. \ref{Fig2}(a) and different 
initial phases $\phi_0\in[0,2\pi]$. Analytical result from Eq. (\ref{eq8}) is 
plotted as continuous line. The overlap of the curves is remarkable good. 
Figure \ref{Fig2}(b) shows the overlap of the curves plotted in (a) onto a 
single and hence universal curve. The scaling transformations are: (i) 
$I_{rms}\rightarrow I_{rms}/(\epsilon^{\alpha_1}\gamma^{\alpha_2})$; (ii) 
$n\rightarrow n/(\epsilon^{z_1}\gamma^{z_2})$. The inset of Fig. \ref{Fig2}(b) 
shows the exponential decay as predicted by Eq. (\ref{eq11}). The control 
parameters used in the inset were $\epsilon=10^2$ and $\gamma=10^{-5}$ and 
with the initial action $I_0=10^5$. The slope of the exponential decay obtained 
numerically is $a=9.195874(1)\times 10^{-6}$, which is close to 
$\gamma(2-\gamma)/2\cong9.99995\times 10^{-6}$.

Let us now show applicability of the formalism developed to a far more 
complicate system, indeed a time dependent billiard \cite{Paper15}. The 
boundary confining an ensemble of non interacting particle is written as 
$R(\theta,\eta,t)=1+\eta f(t)\cos(p\theta)$ with $p$ integer. The case of 
$\eta=0$ corresponds to the circle billiard, which is integrable and that has 
foliated phase space \cite{Book5}. For $\eta\ne0$ and $f(t)=const.$ the phase 
space is of mixed kind exhibiting chaos, invariant spanning curves and periodic 
islands \cite{Paper16}. Fermi acceleration \cite{Paper2} is observed when 
$f(t)=1+\epsilon\cos(\omega t)$, where scaling properties \cite{Paper18} are 
also observed. We shall consider that $f(t)=1+\epsilon\cos(\omega t+Z)$ where 
$Z\in[0,2\pi]$ is a random number generated at each collision of the particle 
with the moving boundary. The dynamics of each particle is given in terms of a 
$4-D$ nonlinear mapping for the variable velocity of the particle $V_n$, 
instant of the collision $t_n$, polar angle $\theta_n$ and angle that the 
trajectory of the particle makes $\alpha_n$ with a tangent line at the instant 
of the collision. The velocity of the boundary at the instant of the impact is 
$\vec{V}_b(t)={{d\vec{R}_b}\over{dt}}(t+Z)$. The reflection laws are given by 
$\vec{V}^{\prime}_{n+1}\cdot\vec{T}_{n+1}=\vec{V}^{\prime}_{n}\cdot\vec{T}_{n+1}
$ and 
$\vec{V}^{\prime}_{n+1}\cdot\vec{N}_{n+1}=-\gamma\vec{V}^{\prime}_{n}\cdot\vec{N
} _ { n+1 }$ where $\gamma\in[0,1]$ corresponding to a restitution coefficient. 
The case of $\gamma=1$ leads to a non dissipative case while $0<\gamma<1$ 
corresponds to the dissipative case. $\vec{T}$ and $\vec{N}$ are the tangent 
and normal unit vectors at the instant of the impact and $^{\prime}$ is to 
consider the momentum conservation law at the moving referential frame. The case 
$\gamma=1$ leads to unlimited diffusion for the velocity of the particles, 
hence producing Fermi acceleration while $\gamma<1$ suppress such a diffusion 
generating a set of points in the phase far from the infinity. The diffusion 
coefficient obtained for this model is
\begin{equation}
D(\eta,\epsilon,\gamma,n)={{\overline{V^2}_n}\over{4}}+{{
(1+\gamma)^2\eta^2\epsilon^2}\over{16}},
\label{eq12}
\end{equation}
where the expression for $\overline{V^2}_n$ is
\begin{equation}
\overline{V^2}(n)=V_0^2e^{-n{{(\gamma^2-1)}\over{2}}}+{{
(1+\gamma)\eta^2\epsilon^2 } \over { 4(1-\gamma) } 
}\left(1-e^{-{{n(\gamma^2-1)}\over{2}}}\right).
\label{eq13}
\end{equation}
As discussed in Ref. \cite{Paper19}, the behavior of the $V_{rms}(n)$ can be 
summarized as: (i) for short $n$, $V_{rms}(n)\propto n^{\beta}$; (ii) for large 
enough $n$, it is observed that 
$V_{sat}\propto(1-\gamma)^{\alpha_1}(\eta\epsilon)^{\alpha_2}$, (iii) finally 
the crossover iteration number is written as $n_x\propto 
(1-\gamma)^{z_1}(\eta\epsilon)^{z_2}$. Doing the same procedure we made along 
on the paper we end up with the following set of critical exponents 
$\alpha_1=-0.5$, $z_1=-1$, $\alpha_2=1$, $z_2=0$ and $\beta=0.5$ as earlier 
obtained in Ref. \cite{Paper19} by using the thermodynamical approach.

As a summary, the diffusion equation is used with a great success to describe a 
transition from limited to unlimited diffusion leading to an analytical 
explanation of the scaling invariance present in such a transition. A set of 
critical exponents, so far obtained from numerical and phenomenological way 
were obtained analytically corroborating for a robust and general interest of 
the procedure.

EDL thanks support from CNPq (301318/2019-0) and FAPESP (2019/14038-6), 
Brazilian agencies. CMK thanks to CAPES for support.

\end{document}